# Global risk minimization in financial markets

Andreas Martin Lisewski

**Department of Molecular and Human Genetics**
**Baylor College of Medicine, Houston, TX 77030, USA**
**Email: lisewski@bcm.edu**
**Phone: +1 713 798 9036**
**Fax: +1 713 798 1116**

**Abstract**

**Recurring international financial crises have adverse socioeconomic effects and demand novel regulatory instruments or strategies for risk management and market stabilization. However, the complex web of market interactions often impedes rational decisions that would absolutely minimize the risk. Here we show that, for any given expected return, investors can overcome this complexity and globally minimize their financial risk in portfolio selection models, which is mathematically equivalent to computing the ground state of spin glass models in physics, provided the margin requirement remains below a critical, empirically measurable value. For markets with centrally regulated margin requirements, this result suggests a potentially stabilizing intervention strategy.**

Large and abnormal fluctuations in financial markets can spread into other parts of the global economy with unwanted and often incalculable effects—as has been observed drastically in recent times. To contain risk and to avoid volatility spillover, a key priority is to minimize risk in today's volatility spreading financial markets (*1, 2*). Important examples of such market places include exchanges where stocks, commodities, futures and other financial products can be bought and sold short by using leverage on margin accounts held by investors. A main financial decision problem in these markets is, for a given expected return $r_P$, to distribute the available capital among multiple assets, which comprise a portfolio $P$ of size $n$, so to minimize the overall risk.

In modern portfolio selection models this goal can be mathematically formulated as finding the global minimum of a risk function (*4-7*),

$R = 1/2 \sum_{i,k=1}^{n} C_{ik} p_i p_k - \sum_{i=1}^{n} p_i r_i - \gamma \sum_{i=1}^{n} p_i s_i$, where $p_i$ is the positive or negative amount of capital invested in asset $i$, and $s_i = \text{sign}(p_i) \in \{-1, 1\}$ are binary *spin* variables; $r_i$ is the expected return of asset $i$ such that $r_P = \sum_{i=1}^{n} r_i p_i$; $C_{ik}$ is the covariance between assets $i$ and $k$; and $\gamma$ is the margin account requirement which sets the fraction of capital that the investor must deposit in a margin account before buying or selling short assets. With the inverse $C^{-1}$ of the covariance matrix $C$ the minimum risk distribution $p = (p_1,\ldots, p_n)$ becomes $p_i = \sum_k C_{ik}^{-1} r_k + \gamma \sum_k C_{ik}^{-1} s_k$. It is known that finding the absolute risk minimum is computationally equivalent to the ground state problem of the random field Ising model (*4, 6*). This is evident after inserting $p$ into the risk function while neglecting fixed terms that do not depend on spin variables: $R = -1/2 \sum_{i,k=1}^{n} J_{ik} s_i s_k - \sum_{i=1}^{n} h_i s_i$, where we introduced an interaction term $J_{ik} = \gamma C_{ik}^{-1}$, and a random field $h_i = \sum_{k=1}^{n} C_{ik}^{-1} r_k$. Because covariance between assets can be both positive and negative (see, for example, inset in Fig. 1A), globally minimizing risk means finding a ground state of the random field Ising model with random spin glass interactions, which in general belongs to the class of NP-complete decision problems (*8, 9*) and for which efficient computational algorithms remain unknown. The computational intractability arises from the non-convexity of the cost function $R$; non-convex problems are much harder to solve computationally than convex optimization problems for which efficient algorithms exist (*10*). In the context of financial markets, the non-convexity of the spin glass model prevents equilibration into a ground state and is viewed as an inherent source of risk (*3, 4*).

We can now demonstrate that ground states are efficiently accessible in the random field spin glass Ising model provided the margin requirement $\gamma$ remains below the critical value $\gamma_c = \left[ \max_i \left( \sum_{k=1}^{n} |\Delta_{ik}| \right) \right]^{-1}$, where $\Delta$ is the Laplacian matrix defined as $\Delta = \text{diag}(\sum_{i=1}^{n} C_{ik}^{-1}) - C^{-1}$. This upper bound on the margin requirement ensures that there exists a related but convex risk function

$R_c = 1/2 \sum_{i,k=1}^{n} J_{ik}(s_i - s_k)^2 + \sum_{i=1}^{n} (h_i - s_i)^2$, which in matrix form reads

$R_c = (s-h)^T(s-h) + \gamma s^T \Delta s$. We note that in the special and simpler case with non-negative interactions $J_{ik} \geq 0$ similar objective functions have been studied in semi-supervised machine learning (11). Our prerequisite $\gamma < \gamma_c$ thus makes the Hessian matrix $H_c = 1 + \gamma \Delta$ positive definite such that $R_c$ remains convex with one global minimum even if interaction is described by a random mix of positive and negative numbers, which is the case in the spin glass model. Let $s$ denote this minimum configuration, then $s$ also depicts the ground state $s^*$ of the spin glass Ising model with a random field because assuming the contrary, $R(s) > R(s^*)$, leads to a contradiction. To see this we choose a discrete path of single spin flips that leads from $s$ to $s^*$. At the beginning $R_c(s)$ is a global minimum and nowhere on the path the cost in $R_c$ can be lower. Concurrently, for any spin flip at $i$ the resulting change in $R_c$ equals twice the risk change in $R$, viz. $\Delta R_{c,i} = R_c(-s_i) - R_c(s_i) = 2\Delta R_i$ with

$\Delta R_i = 2s_i(h_i + 1/2 \sum_{k=1}^{n} J_{ik} s_k)$, and so nowhere along the path–including its end–the risk in $R$ can be lower than at the beginning. Therefore, in contradiction to the assumption, it is $R(s^*) \geq R(s)$, which proves that $s$ is a global minimum of the Ising model.

This result directly implies that once $\gamma < \gamma_c$ is satisfied any algorithm that converges to an Ising model local minimum will, due to the underlying convexity of $R_c$, reach the optimum risk in $R$; for example, this may be achieved by solving for all $s_i$ the local stability condition through fixed points of the TAP (*Thouless-Anderson-*

*Palmer*) equation $s_i = \text{sign}\left[h_i + \sum_k J_{ik} s_k\right]$. It also shows that below the critical margin requirement not an exponential number $\sim 2^n$ of equivalent local minima in the risk function arises (*4, 7*), each one giving a different selection of the portfolio, but only one distinguished risk optimum. Thus the portfolio risk model significantly loses complexity and a computationally efficient, rational access to the optimum is opened.

To illustrate this general result with a numerical example we compared risk values from actual stock price data evaluated below and above the critical margin requirement. For the calculation of the covariance matrix *C* we used *end-of-day* (EOD) stock prices included in the *Standard and Poor's* 500 (S&P500) index over ten years, from February 1999 to February 2009. Given *C* and a random input distribution of the local field *h*, the portfolio was optimized efficiently by solving the TAP equation through iteration until a fixed point was reached. Relative risk is the lowest possible risk value (which was a negative number in our example) divided by the estimated risk after optimization. The lowest risk was found through exhaustive search in all spin states; this was computationally feasible due to our choice of a small portfolio size (*n* = 16). Consistent with the theoretical prediction Fig. 1A shows that with margin requirements below $\gamma_c$ the relative risk settled at its global minimum, *i.e.* at the spin glass Ising model ground state. The picture changes for $\gamma > \gamma_c$, where strong fluctuations significantly elevate the risk above the ground state; for instance, at $\gamma \gg \gamma_c$, the average relative risk from the TAP solutions leveled out at ~25% above the optimum.

The price data further allowed us to follow the critical margin requirement as a function of portfolio size *n*. From the definition of $\gamma_c$ we can expect a decline at least inversely proportional to portfolio size, $\gamma_c \sim n^{-1}$, indicating that in larger portfolios efficient risk minimization imposes stricter limitations on margins, which is consistent with the estimated decrease in the critical margin requirement (Fig. 1B, graph EOD1). However, the observed deviation from the expected scaling, with $\gamma_c \sim n^\alpha$ and $\alpha \approx -1.8$ in Fig. 1B, suggests that intermittency effects in price fluctuations may also be important. For portfolio sizes below $n \approx 100$, this downward trend was robust against changes in price sample selection (graph EOD5), and against smoothing of the data (graph EOD1s5).

Based on our result, the efficient minimization of risk may provide a market instrument for curbing volatility if financial products are traded below the critical margin requirement, and if investors and traders rationally optimize their portfolios. The second condition is both desirable and realistic in today's highly computerized markets, although it may have been less realistic in the past when computers were not widespread and therefore complex financial decisions were to a lesser degree rational. But the first condition seems to be in conflict with interests of traders and lenders who, in individual contracts, seek to reduce default risk by increasing margins. From a collective market perspective, however, higher margin requirements may have a destabilizing effect through higher transaction costs, which can drive traders from the market place; this may lead to a lower overall liquidity thus making the market more susceptible to volatility (*12, 13*). Hence, in financial markets where minimum margin requirements are regulated a reduction of risk by lowering margins is conceivable. Historically, the possibility of such a regulatory approach is indirectly supported by the fact that both the 1987 and the 1929 financial market crashes were accompanied by an increase in margin requirements which exacerbated liquidity problems and which might have contributed to rapid downfall (*14, 15*).

**Acknowledgements**

The author thanks Peter McPhee for providing historical stock price data.

**References**

1. International Monetary Fund. *Global financial stability report*. (Washington DC, April 2009).

2. G.M. Caporale, N. Pittis, N. Spagnolo, Volatility transmission and financial crises, *J. Econ. Finance* **30**, 376-391 (2006).

3. J.-P. Bouchaud, The (unfortunate) complexity of the economy, *Physics World* **22**, 28-32 (April 2009).

4. S. Galluccio, J.-P. Bouchaud, M. Potters, Rational decisions, random matrices and spin glasses, *Physica A* **259**, 449 (1998)

5. A. Gabor, I. Kondor, Portfolios with nonlinear constraints and spin glasses, *Physica A* 274, 222-228 (1999).

6. B. Rosenow, P. Gopikrishnan, V. Plerou, H. E. Stanley, Random magnets and correlations of stock price fluctuations, *Physica A* **314**, 762-767 (2002).

7. L. Bongini, M. Degli Esposti, C. Giardina, A. Schianchi, Portfolio optimization with short-selling and spin-glass, *Eur. Phys. J. B* **27**, 263-272 (2002).

8. C. P. Bachas, Computer-intractability of the frustration model of a spin glass, *J. Phys. A* **17**, L709-L712 (1984).

9. S. Istrail, Statistical mechanics, three-dimensionality and NP-completeness, In *Proceedings of the 31st ACM Annual Symposium on the Theory of Computing*, 87-96 (ACM Press, 2000).

10. H. Rieger, Frustrated systems: ground state properties via combinatorial optimization, In *Lecture Notes in Physics* **501**, 122-158 (Springer Verlag, 1998).

11. D. Zhou, O. Bousquet, J. Weston, B. Schölkopf, Learning with local and global consistency, In *Advances in Neural Information Processing Systems* **16**, 321-328 (MIT Press, 2004).

12. V.G. France, L. Kodres, J.T. Moser, A Review of Regulatory mechanisms to control the volatility of prices, *Econ. Perspect.* **18**, 15-28 (1994).

13. V.R. Anshuman, Regulatory measures to curb stock price volatility, *Econ. Polit. Week,* **38**, 677-679 (2003).

14. P. Rappoport, E.N. White, Was the crash of 1929 expected? *Amer. Econ. Rev.* **84**, 271-281 (1994).
15. M.J. Warshawsky, E. Dietrich, The adequacy and consistency of margin requirements in the markets for stocks and derivative products, *Staff Studies from the Board of Governors of the Federal Reserve System* (*U.S.*) **158** (1989).

**Figures**

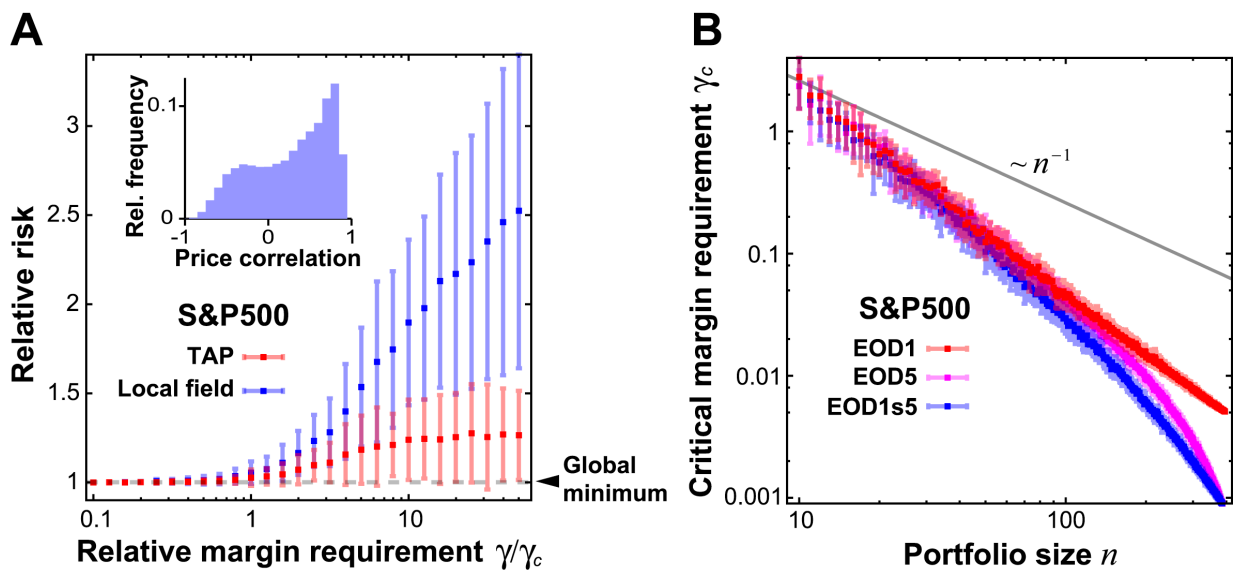

**Figure 1**. (**A**) Portfolio risk can be globally minimized if the relative margin requirement satisfies $\gamma/\gamma_c < 1$. In contrast, for $\gamma/\gamma_c > 1$, the estimated risk undergoes large fluctuations above the optimum. Red data points ("TAP") give the risk from solutions of the TAP equation for $n = 16$ with randomly selected assets from the S&P500 price data, and with a random field $|h_i| \leq 1$. Blue data points ("Local field") depict the risk obtained by taking the sign of local field $h$. Error bars represent standard deviations after 128 random trials. Inset shows the distribution of all price Pearson correlations coefficients between all pairs in the 395 assets taken from the S&P500 index. (**B**) Estimated critical margin

requirement as a function of portfolio size $n$; error bars represent standard deviations from 128 random selections in the S&P500 price data. End-of-day price data selection was done for every trading day available (EOD1) and for every 5 days (EOD5); smoothed price data (EOD1s5) was generated with a sliding boxed-average over 5 trading days.